# Influence of the starting composition on the structural and superconducting properties of $MgB_2$ phase

Y. G. Zhao[1], X. P. Zhang[1], P. T. Qiao[1], H. T. Zhang[1], S. L. Jia[2], B. S. Cao[1]

M. H. Zhu[1], Z. H. Han[1], X. L. Wang[3], B. L. Gu[1]

[1] Department of Physics, Tsinghua University, Beijing 100084, P. R. China

[2] National Lab for Superconductivity, Institute of Physics, Chinese Academy of Sciences, Beijing 100080, P. R. China

[3] Functional Materials Institute, Central Iron & Steel Research Institute, Beijing 100081, P. R. China

We report the preparation of $Mg_{1-x}B_2$ (0=x=0.5) compounds with the nominal compositions. Single phase $MgB_2$ was obtained for x=0 sample. For 0<x=0.5, $MgB_4$ coexists with "$MgB_2$" and the amount of $MgB_4$ increases with x. With the increase of x, the lattice parameter *c* of "$MgB_2$" increases and the lattice parameter *a* decreases, correspondingly $T_c$ of $Mg_{1-x}B_2$ decreases. The results were discussed in terms of the presence of Mg vacancies or B interstitials in the $MgB_2$ structure. This work is helpful to the understanding of the $MgB_2$ films with different $T_c$, as well as the Mg site doping effect for $MgB_2$.

The discovery of superconductivity in the diboride, $MgB_2$, has stimulated worldwide excitement in the scientific community [1]. This binary intermetallic superconductor shows superconductivity at 39 K and its structure is very simple, consisting of alternating layers of Mg atoms and B atoms [1]. Hall effect measurements showed that the charge carriers are holes with carrier density of 1.5?$10^{23}$/$cm^3$ [2]. A significant boron isotope effect was observed in $MgB_2$ [3], suggesting a phonon mediated superconducting mechanism in this compound. It was also shown that this material may be important for the applications [4]. There are some reports about the doping effect of $Mg_{1-x}M_xB_2$ (M are the doping elements) [5-9]. A key issue in these work is whether M atoms are doped on Mg sites. If not, one may get $Mg_{1-x+y}B_2$ (if the doping element reacts with B, then y>0) with the nominal composition. Currently there are a lot of reports on the preparation of $MgB_2$ thin films [10-20]. It was found that $T_c$ of the in-situ prepared films is much more lower than that of the bulk $MgB_2$ samples and it is believed that the low $T_c$ of the in-situ prepared $MgB_2$ films is related to nonstoichiometry of $MgB_2$ and oxidation [13-15, 17, 18]. It was pointed out by Christen et al that the oxidation resulting from residual oxygen in the growth chamber is unlikely to be the leading cause of the lower $T_c$ for the in-situ prepared $MgB_2$ thin films because films prepared with significantly different background pressure are comparable. Anyway the low $T_c$ of the in-situ prepared $MgB_2$ thin films has not been well understood yet. It was also proposed theoretically that $Mg_{1-x}B_2$ (x=0.25) may has higher $T_c$ than that of $MgB_2$ [21]. So it is essential to study the preparation and properties of $Mg_{1-x}B_2$.

In this paper, we report the preparation of $Mg_{1-x}B_2$ (0=x=0.5) compounds with the nominal starting compositions. The results show that single phase $MgB_2$ was obtained for x=0, while $MgB_4$ phase appears and coexists with the "$MgB_2$" phase for 0<x=0.5. With the increase of x, the amount of $MgB_4$ phase increases and the lattice parameters *a* and *c* of the "$MgB_2$" phase decreases and increases respectively. The onset superconducting transition temperature of $Mg_{1-x}B_2$ also decreases with the increase of x. The results were explained by considering the presence of Mg vacancies or B interstitials in the structure of $MgB_2$.

$Mg_{1-x}B_2$ samples with x=0, 0.1, 0.2, 0.3, 0.4, 0.5 were prepared by solid state reaction method. The starting materials were Mg flakes (99.9% purity), and amorphous B powder (99.99% purity) and they are combined in a sealed stainless steel tube in the nominal ratio. The stainless steel tube was then sealed in a quartz ampoule, placed in a box furnace and heated at 950 $^oC$ for two hours, then quenched to room temperature. The phase analysis of the samples was performed using Rigaku D/max-RB x-ray diffractometer with Cu $k_?$ radiation. AC susceptibility of the samples was measured from room temperature to liquid helium temperature.

The x-ray diffraction patterns for x=0 and 0.3 samples are shown in Fig. 1. For x=0 sample, the patterns are consistent with that of $MgB_2$ indicating that the sample is single phase. For $0<x=0.5$ samples, some extra peaks appear besides the peaks of the "$MgB_2$" phase. Careful analysis shows that the extra peaks belong to $MgB_4$ phase. So the "$MgB_2$" phase coexists with the $MgB_4$ phase for $0<x=0.5$ samples. With the increase of x, the amount of the $MgB_4$ phase increases and the amount of the "$MgB_2$" phase decreases for $Mg_{1-x}B_2$ samples. Zi-Kui Liu et al has calculated the temperature-composition phase diagram of the Mg-B system under 1 atmosphere pressure [22]. It shows that for $Mg_{1-x}B_2$ ($0<x<0.5$), which corresponds to the range of atomic fraction of Boron from 0.67 (x=0) to 0.8 (x=0.5) used in their calculation, $MgB_2$ and $MgB_4$ phases coexist below 1550 $^oC$. So our results are consistent with their calculation. With increasing x, the ratio of Mg:B changes from 1:2 to 1:4, thus the amount of $MgB_4$ phase increases and the amount of the $MgB_2$ phase decreases as shown by our results. It was also found that in the x-ray diffraction patterns of $Mg_{1-x}B_2$, the peak positions of the "$MgB_2$" phase also change with x, suggesting that the lattice parameters of the "$MgB_2$" phase change with x. Fig. 2(a) and 2(b) show the variation of the lattice parameter *a* and *c* with x. It can be seen that the lattice parameter *a* decreases and the lattice parameter *c* increases with x. The data for x=0.3 sample shows deviation from the behavior of the data of other samples and this phenomenon was repeated. The reason for this anomaly is not clear at present.

Fig. 3 shows the temperature dependence of AC susceptibility for $Mg_{1-x}B_2$ with x equals to 0 and 0.5 respectively. It clearly shows that the superconducting transition temperature decreases upon increasing x and the amplitude of the ac susceptibility change around the transition also decreases, indicating the volume decrease of the superconducting phase, consistent with the fact that the amount of the $MgB_4$ phase increases and the amount of the "$MgB_2$" phase decreases upon increasing x as shown by the x-ray diffraction results. Shown in fig. 4 is the variation of $T_c$ with x for $Mg_{1-x}B_2$. The $T_c$ value decreases slowly with x for $x=0.2$ samples and then decreases more rapidly for $x>0.2$ samples.

Now we discuss why the lattice parameters and superconducting transition temperature of "$MgB_2$" phase change with x. Currently, the $MgB_2$ thin films prepared by in-situ technique shows much lower $T_c$ than that of the bulk material and the thin films prepared by the ex-situ technique. This is believed to be related to the nonstoichiometry of $MgB_2$ and oxidation [13-15, 17, 18]. Christen et al claimed that the oxidation resulting from residual oxygen in the growth chamber is unlikely to be the leading cause of the lower $T_c$ for the in-situ prepared $MgB_2$ thin films because films prepared with significantly different background pressure are comparable [13]. Based on their experimental results, Shinde et al proposed that the lower $T_c$ of $MgB_2$ thin films may be due to the stabilization of different phase of $MgB_x$ depending upon Mg deficiency [14]. In our experiment, the starting composition is $Mg_{1-x}B_2$, which may favor the formation of $MgB_2$ phase with small amount of Mg vacancies or B interstitials, leading to the change of the lattice parameters and $T_c$ of the "$MgB_2$" phase. For $MgB_2$ thin film growth using pulsed laser deposition (PLD), some nonequilibrium phases (for example, $Mg_{1-x}B_2$ with larger x), which can not exist in the solid state reaction, may form due to the nonequilibrium growth nature of the PLD technique. This has been demonstrated in other compounds [23, 24]. Moreover for $MgB_2$ thin films with lower $T_c$, the lattice parameter *c* is 0.3547 nm [12], which is larger than that of the bulk sample (*c*=0.3521 nm) [12]. This change is consistent with our results. Therefore the above discussion suggests that lower $T_c$ of $MgB_2$ thin films and the decrease of $T_c$ with x for $Mg_{1-x}B_2$ bulk samples in the present work is likely due to the presence of Mg vacancies or B interstitials in the $MgB_2$ phase. Oxidation is unlikely to be the cause of the change of structural and superconducting properties for $Mg_{1-x}B_2$ in our work because the samples were prepared with the same conditions. In fact, if oxidation results in the presence of MgO, it also causes Mg deficiency. The presence of Mg vacancies or B interstitials in the $MgB_2$ phase may result in the change of the lattice parameters of the $MgB_2$ phase and cause the decrease of $T_c$.

In our previous work, we found that Li doping decreases the in-plane B-B distance and the inter-plane B-B distance remains unchanged, and $T_c$ decreases with Li doping [6]. There are some other reports about the doping effect of $Mg_{1-x}M_xB_2$ [5, 8, 25, 26]. For Al doping [5], $T_c$ decreases with the increase of the dopant concentration. Both lattice parameters *a* and *c* decrease with doping, in contrast to the behavior we observed in $Mg_{1-x}Li_xB_2$. For Zn doping [8], it was found that 0.1 Zn can be doped in the $MgB_2$ structure, leading to the increase of both lattice parameter *a* and lattice parameter *c*. A key issue in these doping experiments is whether the doping atoms are doped on Mg sites or not. If not, one may get $Mg_{1-x+y}B_2$ with the nominal composition. Our present work obtained the structural and superconducting properties of $Mg_{1-x}B_2$ and can be used as a reference for the $Mg_{1-x}M_xB_2$ work. For example, the present work strongly supports that Li indeed doped into the $MgB_2$ structure [6] because the behavior of $Mg_{1-x}Li_xB_2$ is quite different from that of $Mg_{1-x}B_2$.

In summary $Mg_{1-x}B_2$ (0=x=0.5) samples were prepared by solid state reaction method. For x=0, single phase $MgB_2$ was obtained. For 0<x=0.5, $MgB_2$ phase coexists with $MgB_4$ and the amount of the $MgB_4$ phase increases with x. With the increase of x, the lattice parameter *c* of the "$MgB_2$" phase increases, while the lattice parameter *a* decreases. Correspondingly the onset superconducting transition temperature of $Mg_{1-x}B_2$ decreases. The results were discussed in terms of the presence of Mg vacancies or B interstitials in $MgB_2$ structure. This work is helpful to the understanding of the $MgB_2$ films with different $T_c$, as well as the doping effect of $Mg_{1-x}M_xB_2$ (M are the doping elements).

## Figure Captions

Fig. 1 x-ray diffraction patterns for $Mg_{1-x}B_2$ with x=0 and 0.3 respectively.

Fig. 2 Variation of the lattice parameters *a* (A) and *c* (B) with x.

Fig. 3 Temperature dependence of AC susceptibility for $Mg_{1-x}B_2$ with x =0 and 0.5 respectively.

Fig.4 Variation of $T_c$ with x for $Mg_{1-x}B_2$ samples.

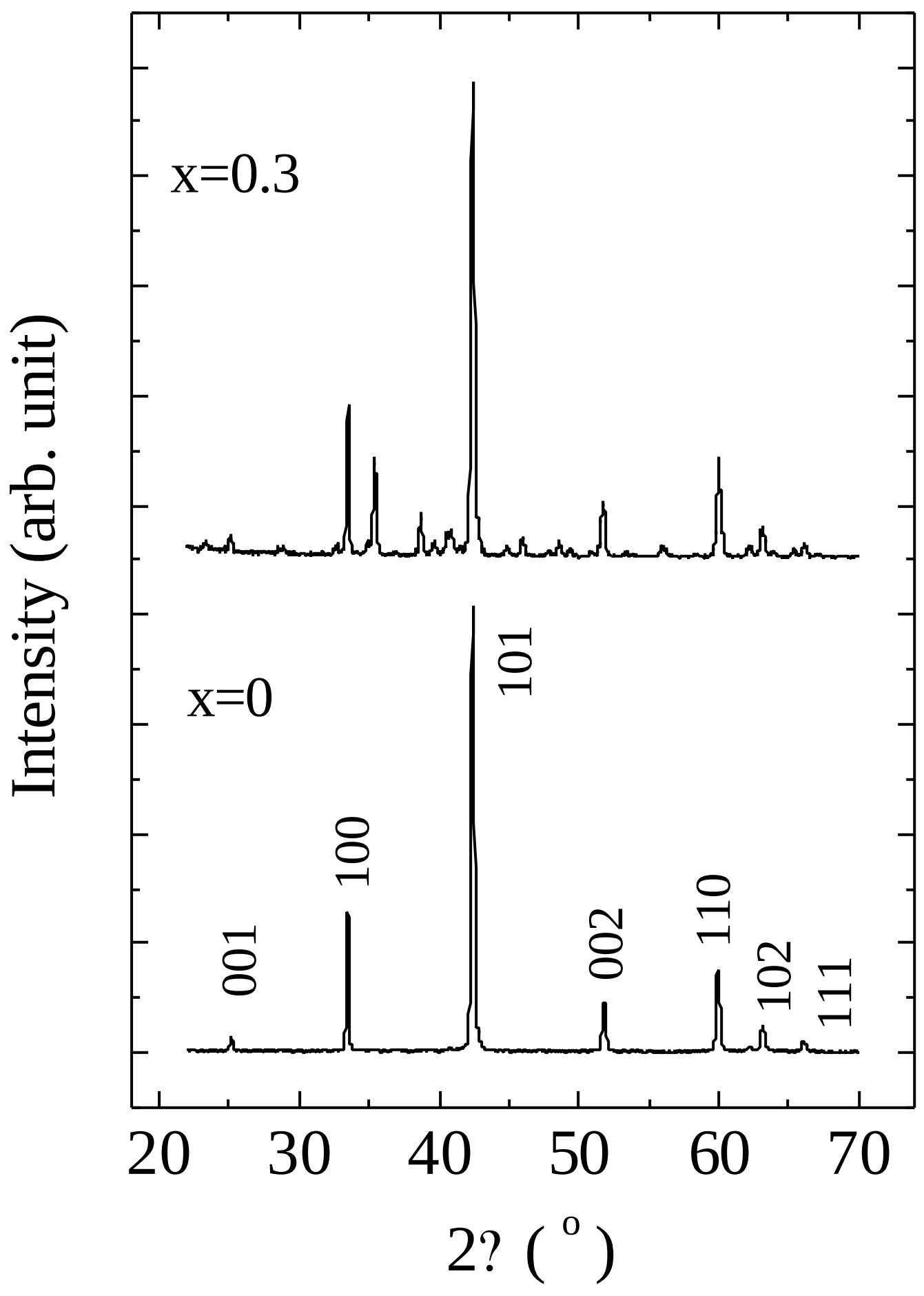

Fig. 1

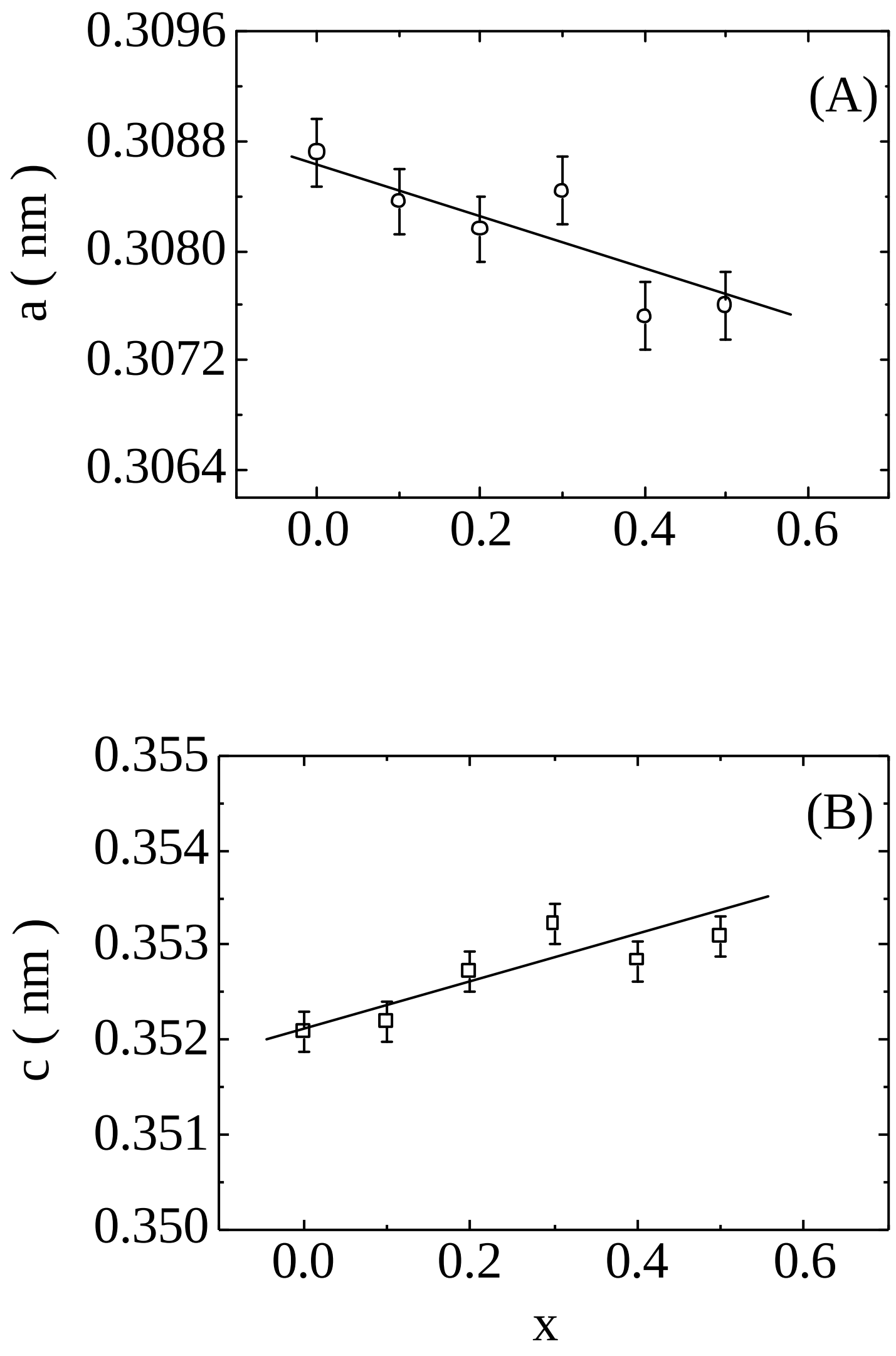

Fig.2

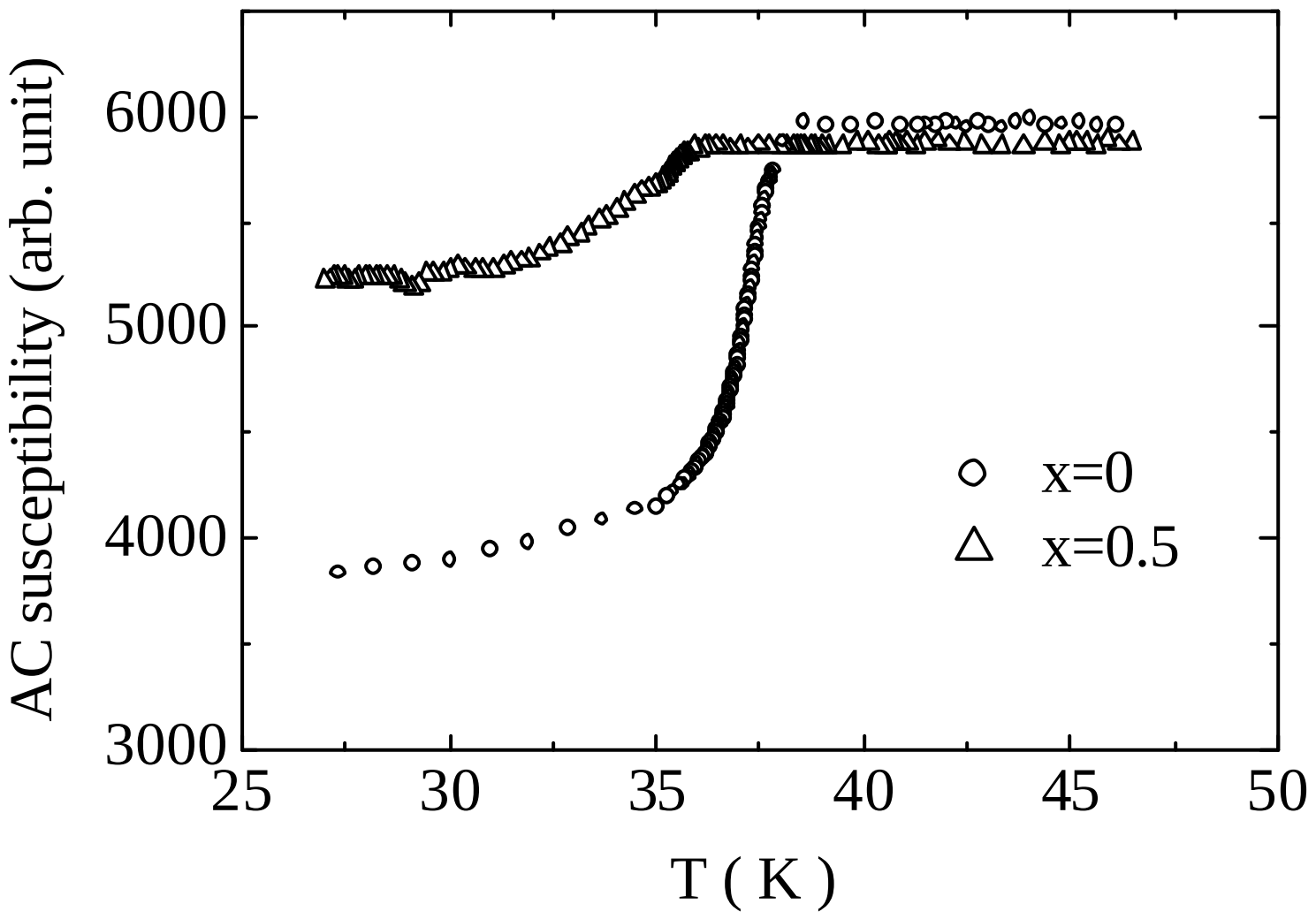

Fig.3

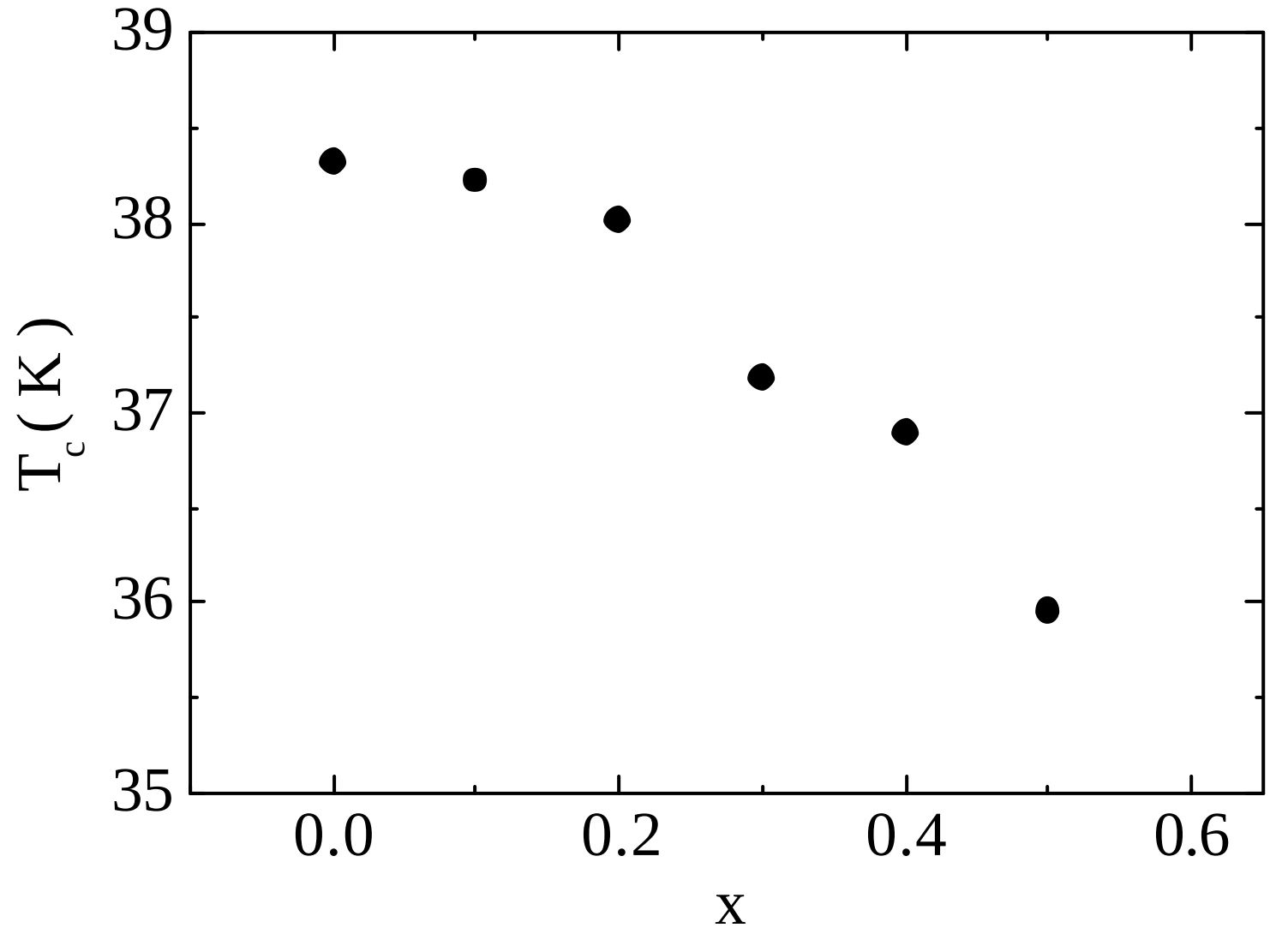

Fig. 4